\documentclass{llncs}

\usepackage{amssymb}
\usepackage{amsmath}
\usepackage{textcomp}
\usepackage{array}
\usepackage{lineno}
\usepackage{color}
\usepackage{comment}

\usepackage{graphicx}

\usepackage{tikz}

\newcommand{\bra}[1]{\langle #1|}
\newcommand{\ket}[1]{|#1\rangle}
\newcommand{\braket}[2]{\langle #1|#2\rangle}
\newcommand{\cent}[0]{\mbox{\textcent}}
\newcommand{\dollar}[0]{\$}

\title{Quantum finite automata: A modern introduction\thanks{Some parts of the material are based on the lectures given by the second author during his visits to Kazan Federal University, Ural Federal University, and Bo\u{g}azi\c{c}i University in 2013.}}

\author{A. C. Cem Say \and Abuzer Yakary{\i}lmaz\thanks{Yakary{\i}lmaz was partially supported by CAPES, ERC Advanced Grant MQC, and FP7 FET project QALGO.}}

\institute{Bo\u{g}azi\c{c}i University, Department of Computer Engineering, Bebek 34342 \.{I}stanbul, Turkey
\and
National Laboratory for Scientific Computing, Petr\'{o}polis, RJ, 25651-075, Brazil
\email{say@boun.edu.tr,abuzer@lncc.br}
}

\authorrunning{A. C. C. Say and A. Yakary{\i}lmaz} 

\begin{document}

\maketitle

\begin{abstract}
We present five examples where quantum finite automata (QFAs) outperform their classical counterparts. This may be useful as a relatively simple technique to introduce quantum computation concepts to computer scientists. We also describe a modern QFA model involving superoperators that is able to simulate all known QFA and classical finite automaton variants. 
\end{abstract}

\section{Introduction}

Due to their relative simplicity, quantum finite automata (QFAs) form a sound pedagogical basis for introducing quantum computation concepts to computer scientists. Early QFA models were problematic, in the sense that they did not embody the full power provided by quantum physics, and led to confusing results where a ``quantum" machine was not able to simulate its classical counterpart. In this paper, we present several simple QFA algorithms which demonstrate the superiority of quantum computation over classical computation. We then systematically construct the definition of a general QFA model, which is able to simulate all known QFA and classical finite automaton variants. 

\section{Preliminaries}\label{sec:prel}

Throughout the paper, $\Sigma$  denotes the input alphabet, not including the left and  right end-markers,  $ \cent $ and $ \dollar $, respectively.  We fix unary and binary alphabets as $ \Sigma = \{a\} $ and $ \Sigma=\{a,b\} $, respectively. A real-time finite automaton does not need to store the input. The given input is fed to the real-time machine from  left to right, symbol by symbol. Moreover, a real-time machine can read $\cent$ before the input and $\dollar$ after the input for pre- and post-processing, respectively. This ability does not increase the computational power of the standard automaton models, but a more detailed analysis is needed for the restricted models. In this paper, our real-time QFA algorithms and models do not use end-markers. Two-way models, on the other hand, have a read-only semi-infinite input tape, composed of infinitely many  cells indexed by the natural numbers, on which the input $w \in \Sigma^* $ is placed as $ \cent w \dollar $ in the cells indexed $0$ to $ |w|+1 $. This tape is scanned by a head which can move one square to the left or right, never moving beyond the end-markers, in each step.

We assume that the reader is familiar with the basics of automata theory. \textit{An $n$-state real-time probabilistic finite automaton} (rtPFA) $M$ is a 5-tuple
\[
	M=(S,\Sigma,\{A_\sigma \mid \sigma \in \Sigma\},s_1,S_a), 
\] 
where $ S= \{s_1,\ldots,s_n\} $ is the set of states, $s_1$ is the initial state, $ S_a \subseteq S $ is the set of accepting states, and $ A_\sigma $ is a left stochastic transition matrix for $\sigma \in \Sigma$ such that $ S_\sigma (i,j) $ is the probability of going from $s_j$ to $s_i$ upon reading $\sigma$. The computation starts in state $s_1$, and the given input is accepted if it finishes in an accepting state. The overall computation on input $w \in \Sigma^*$ can be traced by a stochastic column vector representing the probabilistic distribution of states in each step, whose initial value is  $  v_0 = (1~~0~~\cdots~~0)^T $. After reading the $t$th symbol ($1 \leq t \leq |w|$), the new state vector can be calculated as
\[
 v_t = A_{w_t} v_{t-1}.
\]
The overall acceptance probability of $w$ by $M$ is then
\[
	 f_{M}(w) = \sum_{s_j \in S_a} v_{|w|}(j).
\]
Note that the input is rejected with probability $1 - f_{M}(w)$. If the transition matrices are restricted to contain only zeros or ones as their entries, we obtain \textit{a real-time deterministic finite automaton} (rtDFA).

\section{Basics of quantum computation}\label{sec:quantbasics}

\textit{An $n$-state quantum register} is represented by an $n$-dimensional Hilbert space $ \mathcal{H}_n $ for some positive integer $n$. We denote the standard bases for $\mathcal{H}_n$ as $ \mathcal{B}_n = \{ \ket{q_1},\ldots,\ket{q_n} \}$, where $ \ket{q_j} $ is an $n$-dimensional vector whose $j$th entry is 1, and all other entries are zeros for $ 1 \leq j \leq n $. Each $q$ where $ \ket{q} \in \mathcal{B}_n $ can be seen as a classical state, with the basis state $ \ket{q} $ as its quantum counterpart. We denote the set $\{q_1,\ldots,q_n\}$ by $Q$.

A \textit{(pure) quantum state} of the register is a column vector in $ \mathcal{H}_n $, say,
\[
	\ket{\psi} = \left(  \begin{array}{c} \alpha_1 \\ \vdots \\ \alpha_n	
\end{array}	 \right) = \alpha_1 \ket{q_1} + \cdots + \alpha_n \ket{q_n}, 
\]
which is a linear combination of basis states such that the length of $ \ket{\psi} $ is 1, i.e.
\[
	\sqrt{\braket{\psi}{\psi}} = 1, \mbox{ or equivalently, } | \alpha_1 |^2 + \cdots + | \alpha_n |^2 = 1,
\]
where $ \braket{\cdot}{\cdot} $ is the inner product of any  two given vectors, and, for any $ j \in \{1,\ldots,n\} $,  $ \alpha_j \in \mathbb{C} $ is called the \textit{amplitude} of $ \ket{q_j} $, with $ | \alpha_j |^2 $ representing the probability of being in the $j$th state. 

To observe the classical state of the system, \textit{a measurement in the computational basis}, which determines whether the system is in $\ket{q_1}$, $\ket{q_2}$,$ \ldots $, or $ \ket{q_n} $, is applied. This measurement therefore has $n$ outcomes, respectively ``$1$'',$\ldots$,``$n$''. If the system is in the quantum state  $ \ket{\psi} $ exemplified above before the measurement, the outcome ``$j$'' can be obtained with probability $ p_j = |\alpha_j|^2 $. 

If a system is closed, i.e. there is no interaction (including measurements) with the environment, quantum mechanics dictates that its evolution is governed by some unitary operators. Any operator defined on complex numbers is \textit{unitary} if it is length-preserving, i.e. it maps any quantum state to another quantum state. Thus, we can say that $ \ket{\psi'} = U \ket{\psi} $ is also a quantum state and so its length is 1 too. If $ U \in \mathbb{C}^{n \times n} $ is unitary, then it also has the following equivalent properties: (i) all rows form an orthonormal set, (ii) all columns form an orthonormal set, and  (iii) $ U^{\dagger} U = U U^{\dagger} = I $, where $ U^{\dagger} $ is the conjugate transpose of $ U $. 

One of the earliest quantum finite automaton definitions \cite{MC00,BC01B}  was obtained by ``quantumizing" the rtPFA model of Section \ref{sec:prel} by positing that the transition matrix for each symbol should be  unitary. According to that definition,
\[
	M = \{ Q,\Sigma, \{ U_\sigma \mid \sigma \in \Sigma \}, q_1, Q_a \}
\]  
denotes a real-time quantum finite automaton (rtQFA) with state set $Q$, as described above, and alphabet $\Sigma$. The machine starts out in the quantum state $\ket{q_1}$, which evolves by being multiplied with the unitary matrix $U_\sigma$ whenever the symbol $\sigma$ is consumed, until the end of the left-to-right scanning of the input. At that point, the state is measured, and the input is accepted if any member of the set of accept states $Q_a \subseteq Q$ is observed.

We will see later (Sections \ref{subsec:nonregpoly} and \ref{sec:general}) that one needs somewhat more general operators to reach the full potential of QFAs. But this simple introduction is already sufficient to demonstrate several examples where quantum machines outperform their classical counterparts, as we are going to do in the next section.

\section{Quantum beats classical: Five QFA-based examples}

The algorithms to be  presented in this section are based on a simple common component, which we now describe.

\begin{figure}[!ht]  
  \centerline {\includegraphics[width=0.70\textwidth]{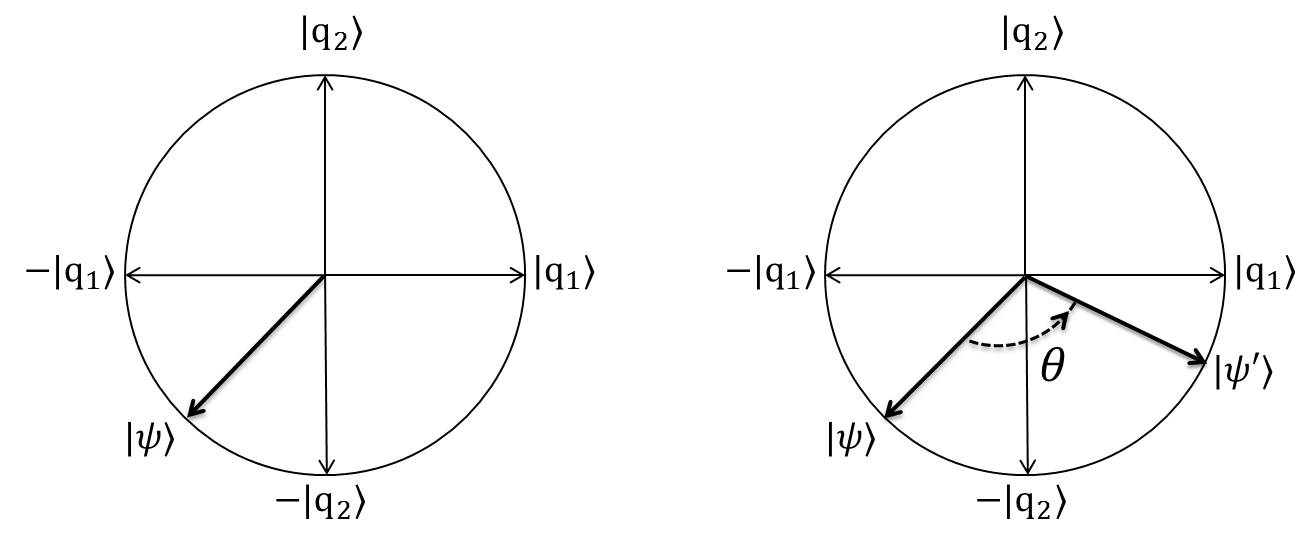}}
  \caption{The representation of rotation $U_\theta$}
\vspace*{-4mm}
\label{fig:image1}
\end{figure}

Consider a QFA whose entire memory can have only two states forming the set $ Q= \{q_1,q_2\} $, i.e. just a \textit{quantum bit (qubit)}. We restrict ourselves to real numbers as amplitudes. Any quantum state of such a single-qubit machine can then be represented as a point on the unit circle of $ \mathbb{R}^2 $, and any possible unitary operator on it is either a reflection or a rotation.  Let $ \theta  $ be the angle of a counterclockwise rotation denoted $ U_\theta $ (see also Figure \ref{fig:image1}):
\[
	U_\theta = \left( \begin{array}{cr}
		\cos \theta &~~ -\sin \theta
		\\
		\sin \theta & \cos \theta
\end{array}	 \right)
~~~~
\mbox{ or }
~~~~
\begin{array}{lcrll}
 U_\theta \ket{q_1} & \rightarrow & \cos \theta \ket{q_1} & + & \sin \theta \ket{q_2}
 \\
 U_\theta \ket{q_1} & \rightarrow & -\sin \theta \ket{q_1} & + & \cos \theta \ket{q_2}
\end{array}.
\]
Note that the $(i,j)$th entry of $U_\theta$ represents the amplitude of the transition from state $ q_j $ to state $q_j$, where $ 1 \leq i,j \leq 2 $. 

It is a well-known fact that if $ \theta $ is a rational multiple of $\pi$, then $ U_\theta $ is periodic, and its repeated application causes the quantum state to visit a finite number of points on the unit circle, returning to the same point after a finite number steps. On the other hand, if $ \theta $ is an irrational multiple of $ \pi $, then $ U_\theta $ is aperiodic and dense on the unit circle, i.e. the quantum state would never visit the same position on the unit circle. 

We proceed with several examples that use such rotations in interesting ways.

\subsection{A QFA can recognize far more tally languages with cutpoint}\label{subsec:moretally}
Define a rtQFA $ R_\theta $ with state set $Q$ as described above, and $ \ket{q_1} $ as the initial state. Our alphabet is unary, $ \Sigma=\{a\} $, and $ R_\theta $ simply applies $ U_\theta $ to the qubit upon reading each $a$. At the end of the computation, the qubit is measured in the computational basis, and the input is accepted if $ q_1 $ is observed. 

It is clear that the empty string is accepted with probability 1. After reading the string $a^k$ ($k>0$), the qubit will be in state
\[
	\ket{\psi_k} = \cos k \theta \ket{q_1} + \sin k \theta \ket{q_2}.
\]
Therefore, the acceptance probability of $ a^k $ by $ R_\theta $ is $ \cos^2 k \theta $.

As can be noticed by the reader, a QFA defines a probability distribution over the strings on its input alphabet, $ \{ (w,f_M(w)) \mid w \in \Sigma^* \} $. So, for the empty string $ \varepsilon $, $ f_{R_{\alpha\pi}} (\varepsilon) $ is always 1 for any $ \alpha \in \mathbb{R} $. If $ \alpha $ is irrational, then there is no nonempty string $ a^k $ such that $ f_{R_{\alpha\pi}}(a^k) $ is 0 or 1. On the other hand, if $ \alpha $ is rational, then there is a minimum positive $ k $ such that $ f_{R_{\alpha\pi}}(a^k)  $ is 1 (and so $ f_{R_{\alpha\pi}}(a^{jk})  $ for any $ j \in \mathbb{N} $). We leave it as an exercise to the reader to determine the values of $ \alpha $ for which $ f_{R_{\alpha\pi}}(a^k)  $ would equal 0.

Since a QFA, say $M$, associates each string with a number in $ [0,1] $, we can split the set of all strings into three groups by picking a cutpoint $ \lambda $ in the interval $ [0,1] $: the strings whose acceptance probabilities are less than, greater than, or equal to the cutpoint. The strings accepted with probability greater than $ \lambda $ form \textit{the language recognized} (or \textit{``defined,"}  in somewhat older terminology) \textit{by $M$ with cutpoint $ \lambda $} \cite{Rab63}:
\[
	L(M,\lambda) = \{ w \in \Sigma^* \mid f_M(w) > \lambda \}.
\]
So, any QFA (or PFA) defines a language with a cutpoint. A language recognized by a PFA with a cutpoint is called \textit{stochastic}, and, it was shown that any language recognized by a QFA with a cutpoint is guaranteed to be stochastic, too \cite{YS11A}. 

In his seminal paper on probabilistic automata, Rabin showed that there are uncountably infinitely many stochastic languages \cite{Rab63}. He presented a 2-state PFA on a binary alphabet, and then showed that  a different language is recognized by that PFA for each different cutpoint. This is not so for tally languages, since 2-state PFAs can define only regular languages, and any $n$-state PFA can define at most $n$ nonregular languages with any cutpoint if the input alphabet is unary \cite{Paz71}. On the other hand, a 2-state QFA can define uncountably infinitely many tally languages \cite{SY14A}, as we argue below:

Let $ U_{\alpha\pi} $ be a rotation with an irrational $ \alpha $, e.g.
\[
	U_{\alpha\pi} = 
	\left( \begin{array}{rr}
		\frac{3}{5} & ~-\frac{4}{5} \\ \\
		\frac{4}{5} & \frac{3}{5}
\end{array}	 \right).
\]
Since $ U_{\alpha\pi}  $ is dense on the unit circle, there is always a $ k $ for any given two different cutpoints $ \lambda_1 $ and $ \lambda_2 $ such that the accepting probability of $ a^k $ lies between $ \lambda_1 $ and $ \lambda_2 $. Thus, $ L(R_{\alpha\pi},\lambda_1) $ and $ L(R_{\alpha\pi},\lambda_2) $ are different. Since there are uncountably many different possible cutpoints, the rtQFA $ R_{\alpha\pi} $ defines uncountably many unary languages. 

\subsection{Nondeterministic QFAs can recognize nonregular languages}\label{subsec:nondet}

\textit{Quantum nondeterminism} is defined as  language recognition with cutpoint 0 \cite{ADH97}. In the classical case, realtime nondeterministic finite automata (equivalently, rtPFAs with cutpoint 0) define only regular languages. On the other hand, rtQFAs with cutpoint 0 can recognize every language in a superset of regular languages known as the exclusive stochastic languages ($\mathsf{S^{\neq}}$) \cite{YS10A}, where  a language is defined to be in $ \mathsf{S^{\neq}} $ if there exists a PFA such that all and only the non-members are accepted with probability $ \frac{1}{2} $. Here, we present a very simple example.

Let $ M $ be a 2-state QFA defined on the binary alphabet $ \Sigma = \{a,b\} $, with initial state $ q_1 $, and $ q_2 $ as the single accept state. After reading an $ a $ (resp., a $ b $), $M$ applies the rotation $ U_{\sqrt{2}\pi} $ (resp., the rotation $ U_{-\sqrt{2}\pi} $). We consider the language recognized by $M$ with cutpoint 0.

It is clear that if $ M $ reads an equal number of $a$'s and $b$'s, the quantum state will be in its initial position $ \ket{q_1} $, and so the accepting probability will be 0. That is, each string containing equal number of $a$'s and $b$'s is definitely not in the recognized language. For any other string, the quantum state ends up on  a point of the unit circle that does not intersect the main axes, and so the acceptance probability will be nonzero, leading to the conclusion that each such string is in the language. Therefore, $M$ recognizes the nonregular language 
\[
	\mathtt{NEQ} = \{ w \mid |w|_a \neq |w|_b \},
\]
where $|w|_\sigma$ denotes the number of occurrences of the symbol $\sigma$ in string $w$, with cutpoint 0 \cite{BC01B}.

\subsection{Succinct exact solution of promise problems}
From a practical point of view, a useful algorithm should classify the input strings with no error, or at least with high probability of correctness. We continue with an exact QFA algorithm.

\textit{A promise problem}  $ P= (P_{yes},P_{no}) $ (defined on $ \Sigma $) is a pair of two disjoint sets $ P_{yes} \subseteq \Sigma^* $ and $ P_{no} \subseteq \Sigma^* $. A promise problem $ P $ is said to be solved by a QFA $ M $ exactly if $ M $ accepts each $ w \in P_{yes} $ with probability 1, and $ M $ accepts each $ w \in P_{no} $ with probability 0. Note that there can be strings outside  $ P_{yes} \cup P_{no} $, and we do not care about the acceptance probabilities of these strings.

Real-time QFAs cannot be more succinct than real-time DFAs in the case of exact language recognition \cite{Kla00}, but things change for certain promise problems \cite{AY12}. For any $k>0$, the promise problem $ \tt EVENODD^k $  is defined as
\[
	\begin{array}{lcl}
		\mathtt{EVENODD^k_{yes}} & = & \{ a^{j2^k} \mid j \mbox{ is a nonnegative even integer} \}
		\\
		\mathtt{EVENODD^k_{no}} & = & \{ a^{j2^k} \mid j \mbox{ is a nonnegative odd integer} \}
	\end{array}.
\]
If we pick $ \theta = \frac{\pi}{2^{k+1}} $, then the rtQFA $ R_{\theta} $ (from Section \ref{subsec:moretally}) can solve $\mathtt{EVENODD^k}$ exactly: It starts in state $ \ket{q_1} $ and, after reading each block of $ a^{2^k} $, it visits $ \ket{q_2}, -\ket{q_1}, -\ket{q_2},\ket{q_1},\cdots   $. So we can solve each $\mathtt{EVENODD^k}$ by a 2-state QFA. On the other hand, any rtDFA solving $\mathtt{EVENODD^k}$ requires at least $ 2^{k+1} $ states \cite{AY12}.\footnote{In fact, any bounded-error PFA or any two-way NFA also requires at least $2^{k+1}$ states for this problem \cite{RY14A,VY14A}.} The interested reader may find it enjoyable to obtain the result for rtDFAs as an exercise. We also refer the reader to the recent works by Gruska and colleagues \cite{GQZ14A,GQZ14B,ZGQ14A} for further results on the succinctness of exact QFAs.

\subsection{Succinct bounded-error language recognition}
Consider the  language
\[
	\mathtt{MOD^p} = \{ a^{jp} \mid j \mbox{ is a nonnegative integer} \}
\]
for some prime number $ p $. Any rtPFA that recognizes $ \tt MOD_p $ with bounded error has at least $ p $ states \cite{AF98}.

\begin{figure}[!ht]  

  \centerline {\includegraphics[width=0.4\textwidth]{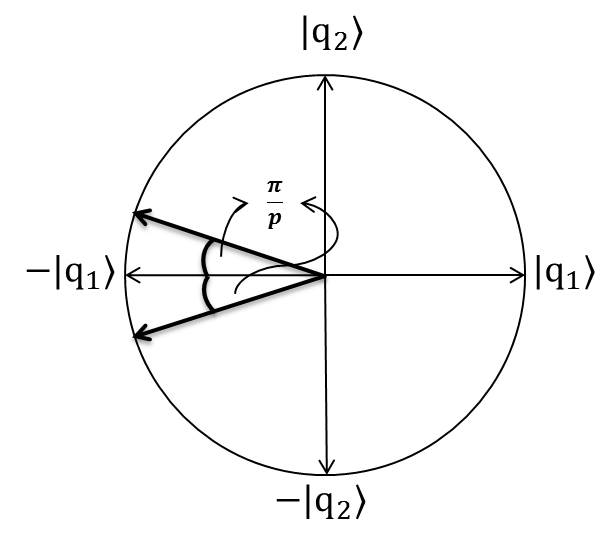}}
  \caption{These two vectors are the closest that the quantum state can get to $-\ket{q_1}$}
\vspace*{-4mm}
\label{fig:image2}
\end{figure}

If we pick a $ \theta = \frac{2\pi}{p} $, the familiar rtQFA $ R_{\theta} $ can accept each member of $ \tt MOD_p $ exactly, and each non-member with  some nonzero probability less than 1. The maximum possible erroneous acceptance probability for non-members is realized for input strings that bring the quantum state closest to $ -\ket{q_1} $ at the end of  its journey on the unit circle, as shown in Figure \ref{fig:image2}. The acceptance probabilities for non-members can therefore be bound by
\[
 \cos^2 \left( \frac{\pi}{p} \right)  = 1 - \sin^2 \left( \frac{\pi}{p} \right), 
\]
and the rejection probability would be at least $ \sin^2 \left( \frac{\pi}{p} \right) $.
As such, the error committed by this family of algorithms nears 1 as $ p $ gets larger. But one can obtain a $ O(\log p) $-state machine for any $ \mathtt{MOD_p} $ for any desired (nonzero) amount of tolerable error by combining several small machines with carefully selected rotation angles. That means that the succinctness gap between QFAs and PFAs can be exponential in the case of bounded-error language recognition \cite{AF98,AN09}. In fact, this bound is tight for the simple rtQFA model, employing only unitary transformations, discussed in this section. We note that any language recognized by an $ n $-state (general) QFA with bounded-error can be recognized by a $ 2^{O(n^2)} $-state DFA, but whether this bound is tight is still an open question \cite{AYXX}.

\subsection{Bounded-error recognition of nonregular languages in polynomial time}\label{subsec:nonregpoly}
Our final example is about two-way automata, which can move their tape head back and forth over the input string, and for which  runtime is therefore an issue. It is known that two-way PFAs cannot recognize any nonregular language with bounded error in polynomial (expected) time \cite{DS90}. We will show how to construct a two-way QFA that recognizes the nonregular language  
\[
	\mathtt{EQ} = \{ w \mid |w|_a = |w|_b \},
\]
with bounded error in polynomial time \cite{AW02}.

Our two-way QFA is actually just a two-way deterministic finite automaton augmented with a qubit (see \cite{AW02} for the general definition). The state set is partitioned to three subsets, namely, the accept, reject, and non-halting states. In each step of the execution, the classical portion of the machine determines either a unitary operator or a measurement in the computational basis to be applied to the quantum register.\footnote{Note that this machine does not fit the simplistic model of Section \ref{sec:quantbasics}, since it allows more than just unitary transformations of the quantum register. See Section \ref{sec:general}.} After this quantum evolution, the machine makes a classical transition based on the scanned input symbol, current classical state, and latest measurement outcome, updating the classical state and head position accordingly. Execution ends when an accept or reject state is entered.

Note that we encountered a quantum machine which recognizes the complement of $\mathtt{EQ}$ with cutpoint 0 in Section \ref{subsec:nondet}. Modifying that machine by setting $q_1$ as a non-halting state and designating $q_2$ as a reject state, we obtain a QFA $M$ that is guaranteed to reject any member of $\mathtt{EQ}$ with probability 0, and to reject non-members with some nonzero probability,  in a single pass of the input from the left to the right.

\begin{figure}[!ht]  
  \centerline {\includegraphics[width=0.5\textwidth]{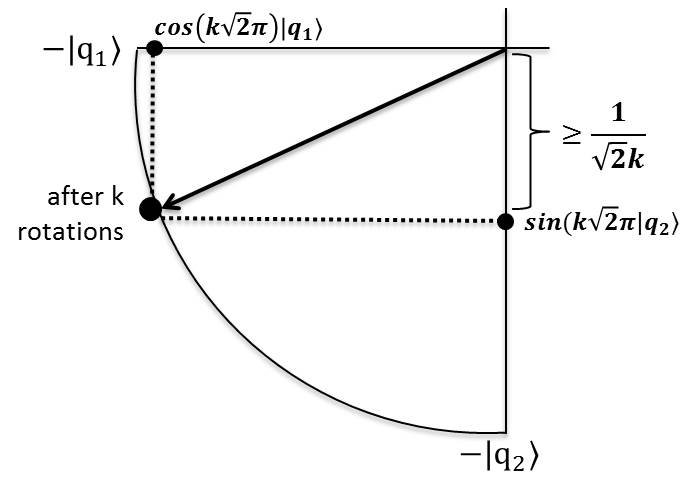}}
  \caption{The minimum distance to the $x$-axis after $k$ rotations (see \cite{AW02} for the proof)}
\vspace*{-4mm}
\label{fig:image3}
\end{figure}

One of the nice properties of the rotation with angle $ \sqrt{2}\pi $ used by $M$ is that, if you start on the $x$-axis ($\ket{q_1}$), the rotating vector always ends up in an orientation that is no closer than an amount proportional to the inverse of the number of rotation steps to the $x$-axis (see Figure \ref{fig:image3}). As indicated in the figure, the rejection probability of any non-member is the square of $ \frac{1}{\sqrt{2} (|w|_a-|w|_b)} $, which can be at least $ p_{rej} = \frac{1}{2|w|^2} $, where $w$ is the input string.

Consider what happens if we augment $M$ to run in a loop, moving its head back to the beginning of the tape and restarting if its left-to-right pass ends in the non-halting state: For input strings in $\mathtt{EQ}$, this new machine would run forever. If the input is not in $\mathtt{EQ}$, however, it would halt with rejection in polynomial expected time.

All that remains is to fix this machine so that it would eventually halt with acceptance, rather than run  forever, with high probability for input strings in $\mathtt{EQ}$, making sure that this fix does not spoil the property of non-members being rejected with high probability. This is achieved by inserting  a call  to a polynomial-time subroutine which accepts the input with probability $ p_{acc} = \frac{p_{rej}}{2}$ at the end of each iteration of the loop.

So our algorithm for $\mathtt{EQ}$ is:

-Run $M$

-Accept with probability $\frac{1}{4|w|^2} $

-If not halted yet, restart.

Since $M$ never rejects  a member of $\mathtt{EQ}$ erroneously, it is clear that this algorithm accepts every member with probability 1. Any non-members would be  rejected with probability at least
\[
	\sum_{j=0}^{\infty} (1-p_{acc}-p_{rej})^{j}(p_{rej})  =  \frac{1}{p_{acc}+p_{rej}} p_{rej}= \frac{2p_{acc}}{3p_{acc}} = \frac{2}{3},
\]
meaning that the probability of erroneous acceptance is at most $ \frac{1}{3} $, that is the error bound. By repeating this procedure $t$ times, and accepting only when all $t$ runs accept,  the error bound can be reduced to  $ \frac{1}{3^t} $. The expected runtime is polynomially bounded, since we made sure that each iteration of the loop has a sufficiently great probability of halting.

\begin{figure}[!ht]  
  \centerline {\includegraphics[width=0.9\textwidth]{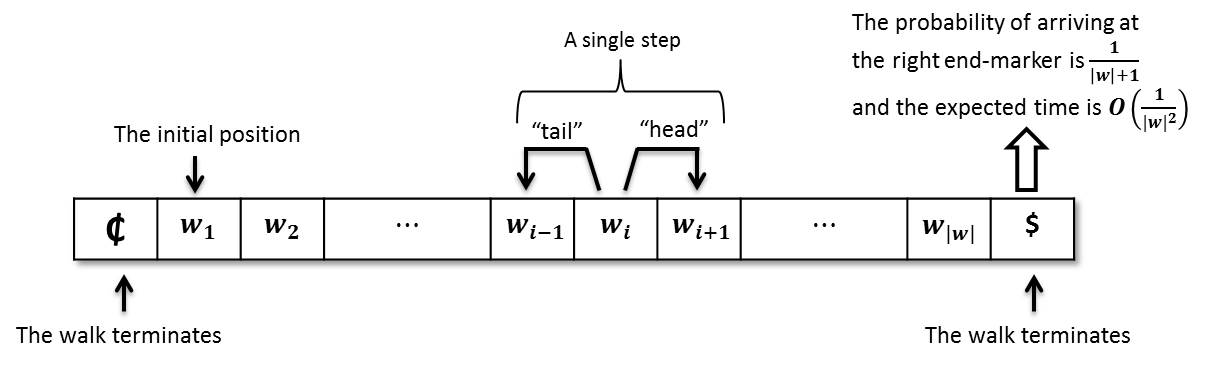}}
  \caption{The details of a random walk on the input $w$}
\vspace*{-4mm}
\label{fig:image4}
\end{figure}

And how do we implement the polynomial-time subroutine that accepts with just the probability described above? This  task is in fact realizable by classical automata. A two-way PFA can easily implement a random walk: The head starts on the first symbol of the input. Then, in each step, a fair coin is flipped, and the head moves to the right (resp.  left) if the result is heads (resp. tails), and, the walk is terminated if the head reaches  an end-marker. The details of such a walk are given in Figure \ref{fig:image4}. A fair coin toss can be obtained by applying a rotation of angle $\frac{\pi}{4}$, i.e.

\[
	\left(  \begin{array}{rr} 
		\frac{1}{\sqrt{2}} & ~-\frac{1}{\sqrt{2}} \\	
		\frac{1}{\sqrt{2}} & \frac{1}{\sqrt{2}}
\end{array}	 \right),
\]
to a qubit in a computational basis state, and then measuring it.

It is another exercise for the reader to show how this subroutine can be designed to accept the input with probability $ p_{acc} = \frac{1}{4|w|^2} $ by using random walks.

\section{General QFAs \label{sec:general}}

As mentioned earlier, the requirement that the program of a QFA should consist wholly of unitary transformations is an overly restrictive one, and several subclasses of regular languages that cannot be recognized by the rtQFA model of Section \ref{sec:quantbasics} have been identified \cite{BC01B}. In fact, this is true even for some proposed generalizations of this QFA model, e.g., \cite{KW97,ABGKMT06,GKK11}. In Section \ref{subsec:nonregpoly}, we saw a two-way QFA model that has classical as well as quantum states, and the classical states govern the computation flow and the determination of whether intermediate measurements or unitary transformations should be performed, depending on both input symbols and previous measurement results. A real-time version of such a model, realizing  a unitary transformation, a projective measurement (see Figure \ref{fig:projective}), and  classical evolution in each step, has been defined formally in  \cite{ZQLG12}, and can easily simulate any rtPFA, for instance. In this section, we focus on a restricted version of this model, and show that the full power of superoperators, generalizing unitary evolution and measurement transformations, is still retained.
\begin{figure}[!ht]
	\fbox{
	\footnotesize
	\begin{minipage}{0.97\textwidth}
	\textit{Projective measurements} are a  generalization  of measurements in the computational basis. Let $Q$ be the set of states, and $ \ket{\psi} $ be the current state. The state set may have been decomposed into some disjoint subsets, e.g. $ Q = Q_1 \cup \cdots \cup Q_k $ for some $ k \in \{1,\ldots,n\} $. Based on this, we can decompose the whole space: 
\[
	\mathcal{H}_n = \mathcal{H}_n^1 \oplus \cdots \oplus \mathcal{H}_n^k, ~~~ \mathcal{H}_n^j = span\{ q \mid q \in Q_j \} ~~~ (1 \leq j \leq n).
\]
Similarly, we can decompose $ \ket{\psi} $ as $ \ket{\widetilde{\psi_1}} + \cdots + \ket{\widetilde{\psi_k}} $ where $ \ket{\widetilde{\psi_j}} \in \mathcal{H}_j $ ($ 1 \leq j \leq n $) and we use the  $\mbox{ }\widetilde{}\mbox{ }$ notation for vectors whose lengths can be less than 1. A measurement operator based on this decomposition  forces the system to collapse into one of these sub-systems when it is applied: There are $ k $ outcomes, say ``$ 1 $'',$\ldots$,``$k$'', and the outcome ``$j$'' can be obtained with probability
\[
	p_j = \sum_{q_l \in Q_j} | \alpha_l |^2 = \braket{\widetilde{\psi_j}}{\widetilde{\psi_j}}~~~(1 \leq j \leq k).
\]
After getting the outcome ``$j$'' ($p_j>0$), the system collapses into the $j$th subspace, and the new state is the normalization of $ \ket{\widetilde{\psi_j}} $, which is $ \frac{\ket{\widetilde{\psi_j}}}{\sqrt{p_j}} $.
	\end{minipage}
	}
	\caption{Projective measurements}
	\label{fig:projective}
\end{figure}

Suppose  that the quantum register of our rtQFA is composed of two systems called the main system (with the set of states $ Q=\{q_1,\ldots,q_n\} $) and the auxiliary system (with the set of states $ \Omega = \{ \omega_1,\ldots,\omega_l \} $, for some $l,n>0$.  So the state space is $ \mathcal{H}_l \otimes \mathcal{H}_n $, and  the set of quantum states is
\[
	\{ (\omega_j,q_k) \mid 1 \leq j \leq l \mbox{ and } 1 \leq k \leq n  \}.
\]
Our machine also has $l$ classical states $\{s_1,s_2,\ldots,s_l\}$, in correspondence with the members of $\Omega$, as will be described below. 

Now suppose that the quantum state is $ \ket{\omega_j} \otimes \ket{\psi}$, where $ \ket{\omega_j}$ is one of the computational basis states of the auxiliary system, and $ \ket{\psi} \in \mathcal{H}_n $. That is, the quantum states of the  auxiliary and main systems are $ \ket{\omega_j} $ and  $ \ket{\psi} $, respectively. It  will be quaranteed that the classical state in this case will be $s_j$, mirroring the auxiliary system state.

 We will trace the execution of our machine for a single computational step. The unitary operator  $ U_{s_j,\sigma} $ to be applied to the quantum register is determined by the classical state $s_j$, and the scanned symbol $ \sigma $. All such operators of this machine are products of two matrices
\[
	U_{s_j,\sigma}= U_\sigma  U_{s_j},
\]
where the functionality of $U_{s_j}$ is to rotate the the auxiliary state to $\omega_1$ from $\omega_j$, so that the operator $ U_\sigma $ finds the quantum state of the overall system to be
\[
	\ket{\Psi}=( \ket{\psi}(1),\ldots,\ket{\psi}(n),\underbrace{0,\ldots,0}_{n \mbox{ times}},\ldots, \underbrace{0,\ldots,0}_{n \mbox{ times}})^\dagger
\] 
before it acts.
 Note that only the first $n$ columns of  $ U_\sigma $  determine the state attained after the evolution. Let us partition $ U_\sigma $ to $ n \times n $ blocks. There are $ l^2 $ of these blocks, but only the ``leftmost" $ l $, designated $E_1$ through $E_l$ below, are significant for our purposes:
\[
	U_\sigma=\left(  \begin{array}{c|c|c|c}
		E_1 & * & \cdots & *
		\\ \hline
		E_2  & * & \cdots & *
		\\ \hline
		\vdots & \vdots & \ddots & \vdots 
		\\ \hline
		E_l  & *& \cdots & *
\end{array}	 \right) .
\] 
The reader can also verify that the state obtained  after applying $ U_\sigma $ to $\ket{\Psi}$ is
\[
 \ket{\Psi'} = 	\left( \begin{array}{c}
 		\ket{\widetilde{\psi_1}} \\
 		\ket{\widetilde{\psi_2}}
 		 \\
 		 \vdots
 		 \\\ket{\widetilde{\psi_l}} 
 		 \end{array} \right),
\]
where $\ket{\widetilde{\psi_i}}= E_i \ket{\psi}$ for $i\in\{1,\cdots,l\}$. Following this evolution, the auxiliary system is measured in the computational basis, which amounts to a projective measurement on the composite system. (This measurement is independent of the input symbol processed at the current step.)  The probability of obtaining outcome ``$k$'' (where $1\le k \le l$) is $ p_k=\braket{\widetilde{\psi_k}}{\widetilde{\psi_k}} $, and if ``$k$'' is observed $ (p_k>0) $, the quantum state of the main system collapses to $ \ket{\psi_k} = \frac{\ket{\widetilde{\psi_k}}}{\sqrt{p_k}} $. As the final action of every computational step for any input symbol, the classical state is set to $ s_k $ to mirror the observation result ``$k$''. 

The reader might have  noticed that all the information relevant to the computation is kept in the main system, and only the first $ l $ columns of the unitary operator actually affect the computation. It is therefore possible to trace the entire computation by just knowing $ \mathcal{E} = \{ E_1,\ldots,E_l \} $, and forgetting about the classical state and the auxiliary system. $ \mathcal{E} $ is in fact what is called a \textit{superoperator}, and each of the $ E_j $ are said to be its \textit{operation elements}. Since they are composed of $l$ orthonormal columns of a unitary operator, the operation elements satisfy the following equation that the reader can prove as an exercise:
\[
	\sum_{j=1}^l E_j^{\dagger} E_j = I.
\]

We can now  focus only on the main system as our machine, and think of the classical state and the auxiliary system as representing the environment that the machine interacts with. In that view, the computational step described above has caused the machine to be in a mixture of pure states, appropriately called a \textit{mixed state}, which can be represented as 
\[
 \{ (p_j,\ket{\psi_j}) \mid 1 \leq j \leq l \}.
\]
But there is a more convenient way to represent such a mixture as a single mathematical object, called a \textit{density matrix}. Here is how to obtain the  density matrix describing the mixture above:
\[
	\rho = \sum_{j=1}^l p_j \ket{\psi_j}\bra{\psi_j} .
\]
$(\bra{\psi_j}$ is defined to be the conjugate transpose of $\ket{\psi_j}$.) The reader can verify that, for each $j$, the $ j $th diagonal entry of $ \rho $ represents the probability of the system being observed in the $j$th state. Therefore, the sum of all diagonal entries, the trace of the matrix ($ Tr(\rho) $), is equal to 1. 

A simple derivation reveals how this mixed state resulted from the pure state $\ket{\psi}$ through the application of our superoperator, as we represent $ \rho $ in terms of $ \ket{\psi} $ and the operation elements:
\[
	\rho = \sum_{j=1}^l p_j \ket{\psi_j}\bra{\psi_j} 
	= \sum_{j=1}^l p_j  \frac{\ket{\widetilde{\psi_j}}}{\sqrt{p_j}} \frac{\bra{\widetilde{\psi_j}}}{\sqrt{p_j}} 
	= \sum_{j=1}^l \ket{\widetilde{\psi_j}} \bra{\widetilde{\psi_j}}
	=  \sum_{j=1}^l E_j \ket{\psi} \bra{\psi} E_j^\dagger.
\]
In general, this is how you apply a superoperator to a state $\rho$ to obtain the new state $\rho'$:
\[
	\rho' = \mathcal{E} (\rho) =  \sum_{j=1}^l E_j \rho E_j^\dagger.
\]
A density matrix $\rho$ has the following properties:
(i) $ Tr(\rho) =1 $, (ii) it is Hermitian, and (iii) it is semi-positive.
Moreover, any density matrix corresponds to an actual mixed state.

We are ready to give the formal definition of a general QFA \cite{Hir10,YS11A}. An $n$-state QFA  $\mathcal{M}$ is a five-tuple
\[
	 \{ Q,\Sigma, \{ \mathcal{E}_\sigma \mid \sigma \in \Sigma \}, q_1, Q_a \},
\]  
where (i) $ Q = \{ q_1,\ldots,q_n \} $ is the set of states, $ q_1 \in Q $ is the initial state, and $ Q_a \subseteq Q $ is the set of accepting states; (ii) $ \Sigma $ is the alphabet; and, (iii) $ \mathcal{E}_\sigma $ is the superoperator defined for $ \sigma \in \Sigma $ with $ l_\sigma $ operation elements: $ \{ E_{\sigma,1},\ldots,E_{\sigma,l_\sigma} \} $.

Let $ w \in \Sigma^* $ be the input. The computation starts in state $ \rho_0 = \ket{q_1}\bra{q_1} $. After reading each symbol, the defined superoperator is applied,
\[
	\rho_t = \mathcal{E}_{w_t} (\rho_{t-1}) = \sum_{j=1}^{l_\sigma} E_{w_t,j} \rho_{t-1} E_{w_t,j}^\dagger,
\]
where $ 1 \leq t \leq |w| $. After reading the whole input, a measurement in the computational basis is made, and the input is accepted if one of the accepting states is observed. The overall accepting probability can be calculated as 
\[
	f_{M} (w) = \sum_{q_j \in Q_a} \rho(j,j).
\]

\noindent
\textbf{Simulation of classical machines:} Let $ v $ be the state of  an $n$-state probabilistic system, say $P$:
\[
	\left(  \begin{array}{c}
		p_1 \\ p_2 \\ \vdots \\ p_n
\end{array}	 \right),
~~~ \sum_{j=1}^n p_i = 1.
\]
An $n$-state quantum system, say $M$, can represent $v$ as
\[
	\ket{v} = 
	\left(  \begin{array}{c}
		\sqrt{p_1} \\\sqrt{p_2} \\ \vdots \\ \sqrt{p_n}
\end{array}	 \right).
\]
Suppose that $P$ is updated by a stochastic matrix $ A $, i.e. $ v'=Av $. Let us focus on the $j$th state, whose probability is $ p_j $ in $v$. Operator $A$ maps $p_j$ to
\[
	p_j \left(  \begin{array}{c}
			A(1,j) \\ A(2,j) \\ \vdots \\ A(n,j)
\end{array}	 \right),
\]
that represents the contribution of the $j$th state of $ v $ to $v'$. Now, we define a superoperator $\mathcal{E}$ with $n$ operation elements $  \{E_1,\ldots,E_n\}  $ that simulates the effect of $A$ as follows: The $j$th column of $E_j$ is $ (\sqrt{A_{1,j}},\sqrt{A_{2,j}},\ldots,\sqrt{A_{n,j}})^T $, and all other entries are zeros. (The reader can easily verify that $\mathcal{E}$ is a valid superoperator.) Then $E_j$ maps $\ket{v}$ to 
\[
	\sqrt{p_j} \left(  \begin{array}{c}
			\sqrt{A(1,j)} \\\sqrt{A(2,j)} \\ \vdots \\ \sqrt{A(n,j)}
\end{array}	 \right),
\] 
that reflects the contribution of the $j$th column of $A$. By considering all operation elements, we can follow that the whole effect of $A$ on $v$ can be simulated by $\mathcal{E}$. Therefore, the evolution of  $P$ can be simulated by $M$ by using a corresponding superoperator for each stochastic operator if a measurement in the computational basis is applied at the end of the computation of $\mathcal{M}$. 

A straightforward conclusion is that any rtPFA can be simulated by a rtQFA having the same number of states. 
Moreover, since the tensor product of two superoperators is another superoperator, a rtQFA can simulate the computations of two rtQFAs in parallel. 
Therefore, rtQFAs are sufficiently general to simulate all known classical and quantum real-time finite state automata.\footnote{We refer the reader to \cite{BMP03A,LQZLWM12,ZQLG12} as some examples of classically enhanced rtQFAs .}

For two recent surveys on QFAs, we refer the reader to \cite{QLMG12} and \cite{AYXX}.

~\\
\noindent
\textbf{Acknowledgement.} We first met quantum finite automata in Prof. Gruska's book on quantum computing \cite{Gr99}, for which we would like to extend him our thanks.

\bibliographystyle{plain}
\bibliography{tcs}

\end{document}